\documentstyle[12pt]{article}
\def\beq{\begin{equation}}   \def\eeq{\end{equation}}
\begin{document}
\begin{titlepage}

\begin{flushright}
NYU-TH-99/05/01\\
\end{flushright}

\vspace{0.3cm}

\begin{center}
\baselineskip25pt

{\Large\bf  Infrared Hierarchy, Thermal Brane Inflation and
Superstrings as Superheavy Dark Matter}

\end{center}

\vspace{0.3cm}

\begin{center}
\baselineskip12pt

{\large Gia Dvali\footnote{Also ICTP, Trieste, Italy}}

\vspace{0.2cm}
Physics Department, New York University, New York, NY 10003

\vspace{2.5cm}

{\large\bf Abstract} 

\vspace*{.25cm}

\end{center}
In theories with TeV scale quantum gravity the standard model particles
live on a brane propagating in large extra dimensions. Branes
may be stabilized at large (sub-millimeter)
distances from each other, either due to weak Van der Waals type
interactions, or due to an infrared analog of Witten's inverse hierarchy
scenario. In particular, this infrared stabilization may be responsible
for a large size of extra dimensions. In either case,
thermal effects can drive a brief period of the late inflation
necessary to avoid the problems
with high reheating
temperature and the stable unwanted relics. The main reason is that
the branes
which repel each other at zero temperature can be temporarily glued
together by thermal effects. It is crucial that
the temperature needed to stabilize branes on top of each other can be
much smaller than the potential energy of the bound-state, which drives
inflation. After 10-15 $e$-foldings bound-states cool below the critical
temperature
and decay ending inflation. The parallel brane worlds get separated
at this stage and superstrings (of a sub-millimeter size) get stretched
between them. These strings can have the right density in order to serve
as a superheavy dark matter.
\end{titlepage}

\section{Introduction.}
It was suggested recently, that the fundamental scale of quantum gravity
may be as low as TeV, provided there are $N$ large new dimensions
to which gravity can propagate\cite{add}.
The relation between the observed Planck scale $M_P$ and the
fundamental one $M$ is then given by
\begin{equation}
M_P^2 = M^{N + 2}V_N, \label{planckscale}
\end{equation}
where $V_N \sim R^N$ is the transverse volume of extra
space. In this picture,
all the standard model particles must live in a brane (or a set of branes)
with $3$ extended space dimensions.
\footnote{The attempt of lowering the Planck scale to $M_{GUT}\sim
10^{16}$GeV goes back to\cite{witten}. In a different
context lowering the string scale to TeV, without lowering
the fundamental Planck scale was suggested in \cite{joe}.
Dynamical localization of the fields
on a (solitonic) brane embedded in a higher dimensional universe has been
studied earlier in the field theoretic context\cite{localization1/2},
\cite{localization1}}.
The supersymmetry breaking in the observable world may then result from
a non-BPS nature of our brane Universe\cite{dshifman}.

Perhaps the most natural realization of this picture is
via the $D$-brane constructions (see\cite{polchinski} for an introduction).
The standard model fields can be identified with the open string modes
stuck on a $D$-brane, whereas gravity comes from the closed
string sector propagating in the bulk \cite{aadd,st, bw}.

Cosmological constraints discussed in\cite{add1}, suggest that
in such a scenario there is an absolute bound on the
reheating temperature of the Universe due to the overproduction of
the bulk Kaluza-Klein gravitons. In particular, for $N=2$ this bound gives
$T_R \leq T_* \sim$ MeV, and even for larger $N$, $T_*$ stays well below
TeV. Obviously, with such a low reheating temperature it is
hard to accommodate conventional four-dimensional scenarios
for inflation or baryogenesis\cite{bdav}.

However, brane picture opens up the new, intrinsically
high-dimensional
mechanisms, both for inflation\cite{braneinflation} and
for baryogenesis\cite{baryo}. Baryon asymmetry for instance, may result
because of the inflow of the baryonic charge to our brane world, due to
creation of the baby branes, or due to baryon number transport in the 
brane-brane collision\cite{baryo}.
Such a scenario of the baryogenesis does not necessarily require high
reheating temperature. In fact, the temperature of the inflaton decay
products, after they thermalize, can be lower that the typical baryon mass.

Recently a new inflationary mechanism, brane inflation, has been proposed
\cite{braneinflation}. Inflation is driven by the displaced branes that
slowly fall on top of each other. This slow-fall in due to the weak
inter-brane attraction
and translates as slow-roll of the inflaton in an effective
four-dimensional field theory language. Inflation ends by the brane
"collision" which reheats the Universe.
In its most straightforward version this scenario may suffer from a
high reheating temperature problem ($T_R \sim M$), as in particular was
pointed out by Banks, Dine and Nelson\cite{bdn}.

In the present paper we will argue that branes may provide
a built-in mechanism for avoiding this problem. 
In fact, the same high reheat temperature
can trigger a secondary stage of the brief brane inflation with a
very low reheating temperature.

We show that this scenario can have an interesting
byproduct, producing superstrings (of sub-millimeter length), which can
serve as a superheavy dark
matter!

 Our scenario can be summarized as follows. 
In the absence of supersymmetry, branes are not BPS states and
experience Van der Waals type interactions due to exchange of the
light bulk modes (e.g. graviton, dilaton and Ramond-Ramond (RR) fields).
Depending on the balance between the repulsive and attractive
messenger forces, the branes can repel each other at the short
distances and get stabilized at some large distance from one another.
Alternatively, a large distance stabilization can be achieved via
the analog of Witten's inverted hierarchy scenario\cite{wittenh}.
This way of generating the large inter-brane separation
can provide an explanation for the large size of extra dimensions,
without any reference to big input quantities!
(The alternative way would be to postulate some big conserved
numbers, e.g. such as the number of branes, or a topological charge of the
Universe \cite{adm,adm1}).

However, whatever source stabilizes branes at zero temperature, the
thermal effects change the picture.
At high temperature, some branes that would normally repel
each other at zero temperature, get stabilized on top of each other.
In the other words at high $T$ coincident branes correspond to a
meta-stable minimum of the free energy.

The crucial point is that the temperature required for
their stabilization can be much smaller than the potential energy of the
boundstate. This energy drives a brief period of inflation,
with a very low reheating temperature, just
enough to get rid of unwanted relics. When the temperature drops to a
certain
critical value $T_c$, the boundstate becomes unstable and branes roll
away marking the end of inflation. During this process strings get
stretched between branes, which can serve as a new source of a superheavy
dark matter. After rolling away brane-brane bound system oscillates around
the equilibrium point and reheats the Universe to an
acceptably low temperature.

Before proceeding, we want to note that density fluctuations will not be
discussed in the present paper. We'll assume that these fluctuations are
created at the earlier stage (e.g. by an earlier radius inflation
\cite{rad,rad2}).

\section{Van der Waals Forces Between the Branes}

Let us consider the two parallel branes in a space of $N>1$ transverse
dimensions.
Assume that the distance between the branes ($r$) is much larger than
their "thickness", typically given by an inverse scale of tension
$(\sigma)^{1/4} \sim M^{-1}$.
Such branes will then interact via an exchange of the bulk particles. 
Some of these forces (e.g. gravity and dilaton) can give attraction, and
others (e.g. bulk gauge fields) can give repulsion if branes have
the same
sign of charge.

 Sometimes it may happen that, due to supersymmetry, there is a very
precise (in perturbation theory) relation among brane
tension ($\sigma$) and its charge,
so that the resulting force cancels out.
In such a case the two-brane system is a BPS state and there is a zero net
force between them. This is, for instance, true for the parallel $D$-branes.
In string theory picture this can be understood as vanishing
of a tree level amplitude with closed string exchange, or alternatively
of the one-loop open string amplitude stretched between the 
branes.

 In an effective low energy field theory picture this can be
understood as the graviton-dilaton attraction compensated by the RR
repulsion. 

However, in  the real world supersymmetry is broken and we expect that
the force is no longer zero. Then at very short
distances
$r \sim M^{-1}$ the interaction between the
low-energy modes that are
localized on the different branes as well as the string modes that are
stretched between the two become important and the sign of the potential
depends on these couplings. We will assume that the overall interaction
is repulsive. At the distances $r \gg M^{-1}$, however,
these modes decouple and their
contribution dies away exponentially fast. At large distances the brane
interactions are governed by the two sources: 1) Exchange of the
light bulk modes, e.g. such as the graviton, dilaton or RR fields; and
2) by the tension of the strings that are {\it actually} stretched between
the branes. The second source provides a linear (in $r$) confining
potential between the branes. However, its existence depends on
a number density of such strings. It is very hard to produce them while
branes are
already separated. So such strings mostly will be important in the case
when branes initially come close and get separated later.
For the low energy observer these states will look as superheavy
particles of the mass
\begin{equation}
m_{string} \sim M^2r 
\label{thepotential}
\end{equation}
and can play the role of a dark matter. Below we will show that such
states
can be actually produced after the thermal brane inflation.

 With the account of the above two sources, the brane interaction potential
at large distances assumes the form:
\begin{equation}
V(r) = M^4(\alpha + b_i{e^{-m_ir} \over (Mr)^{N - 2}} -
{1\over (Mr)^{N -2}} + kr)
\label{thepotential}
\end{equation}
where $\alpha, b_i$ are model-dependent constants of order one.
$k$ is proportional to the density of the stretched strings per unit
brane-volume. 

In our scenario the linear term even if presented initially
will quickly redshift away during the first stage of inflation,
but can be
recreated
after reheating, since branes sit on top of each other.
So the initial density of the stretched strings will be given by the 
reheating
temperature after the first inflation $T_{in}$ (see below).

 So to study a zero-temperature (zero particle density) behavior of
branes, one can ignore the contribution of the confining energy and
concentrate on the first term. The Yukawa type potential comes from the
masses of the bulk modes and the inverse-power term comes from
the gravitational interaction (for $N = 2$, ln$(r)$ behavior should be
understood). If some of the bulk gauge fields are
lighter then $M$, then at the distances $M^{-1} \gg r$ the attractive
gravitational interaction can be dominated by a repulsive gauge
interaction and the over-all potential can be repulsive at the short
distances. Thus, branes can experience the Van der Waals type interactions
at zero $T$.
In particular, this will be the case if the dilaton
is "projected out" from the low energy physics. That is, if it gets a mass
$m_{dilaton} \sim M$ due to some non-perturbative strong dynamics.
Let the mass of the remaining repulsive mode be
$m$ and for simplicity assume the only one such a mode. Then there is a
stable equilibrium point at $r_0 \sim 1/m$ with a zero net force.
So at zero temperature branes will be stabilized at this point.

How small can $m$ be? In the absence of a concrete model $m$ can be
naturally as small as the supersymmetry-breaking scale in the bulk, which
if the SUSY-breaking occurs on a separate distant brane, can be all the
way down to
an inverse millimeter $\sim 10^{-3}$eV\cite{aadd}. In what
follows, we will keep $m$ as a free parameter $M \gg m$ and fix its
value from cosmological constraints.

When the branes are at distance $r_0$ the effective four dimensional
cosmological constant is given by
\begin{equation}
\Lambda_{eff} = V(r_0) + \Lambda_{bulk}V_N 
\label{lambda}
\end{equation}
and must be canceled. This is the usual fine tuning
problem on which we have nothing new to say. However,
if the branes are brought at the distances $r \rightarrow 0$ there will be
an excess of the potential energy resulting into an effective cosmological
term
\begin{equation}
\Lambda_{eff} = V(0) - V(r_0) \sim M^4
\end{equation}
At zero temperature, however, the repulsive potential at $r=0$
is steep. Thus, the branes will quickly relax towards $r_0$ and
no inflation will result.
Below, we will show that the situation can change dramatically when
temperature
effects are taken into
account\footnote{The Van der Waals form of the potential can allow
for the low reheating temperature even for the slow-roll brane
inflation\cite{savasnima}. Here we will be interested only in the
inflation driven by thermal effects.}

\section{Brane Stabilization from Inverted Hierarchy}

From above discussion, it seems not unnatural to expect the Van der Waals
type forces between the branes. The positions of the minima
are then defined by the masses of the bulk fields, and the large
distance
stabilization would require some of these masses to be small.
As said above, this smallness can be due to the smallness of the
supersymmetry breaking scale in the bulk, which can be suppressed
by a bulk volume factor $\sim 1/(MR)^N$. Stated in this way the issue
becomes linked with a largeness of the extra dimensions.
It is natural to ask whether branes can be stabilized at large distances,
without help of the small parameters? This question has an independent
motivation, since in such a case one can invert the issue and try to
explain largeness of the radius by the large inter-brane separation.

In '81 Witten suggested a way generating a large mass scale from
a small one as a result of the dimensional transmutation\cite{wittenh}.
From the first glance the issue is not quite the same, since we
are willing to generate a small mass scale instead of large.
However, the duality between the infrared gravity and ultraviolet
gauge dynamics suggests that the solution may work
also here.

 In the gauge theory description the brane separation
is a VEV of the Higgs field, that gives masses to the open string modes.
Thus, in this picture generation of large inter-brane distance is
equivalent to generation of the large mass scale. Details will be
presented in \cite{zura}, here we will briefly discuss the main idea
relevant for our purposes.

To illustrate the point, consider a toy $N = 4$ supersymmetric
example with a set of $n$ $D3$-branes in space with two large
transverse dimensions. When branes are coincident there is an enhanced
$SU(n)$ symmetry, which gets
broken if some of them get displaced. In supersymmetric limit, the
moduli space can be parameterized by an adjoint VEV $\Sigma$.
Now imagine that some strong dynamics generates a superpotential
for the lowest massless modes
\begin{equation}
W = \phi\Lambda^2 + \lambda\Sigma^3 +...
\label{superpotential}
\end{equation}
where $\Lambda \sim M$ is some typical mass scale of this dynamics
and $\phi^2 = Tr\Sigma^2$.
At the tree level, this leaves $N = 1$ supersymmetry among the
massless modes, but we will assume that heavy modes do not
experience breaking at this stage. What is the moduli space of this
theory? If $n$ is even, for $\phi >> M$ there is a plateau
along which supersymmetry is broken and the gauge symmetry is broken 
to $SU(n/2)\otimes SU(n/2)\otimes U(1)$
at the scale $\phi = M^2r$, (where $r$ is the brane separation).
This means that at the tree level there is a net zero force between the
branes if two sets of coincident $n/2$ branes are displaced.
\footnote{As it stands, the superpotential in Eq(\ref{superpotential}) also
admits the isolated supersymmetric minima with two sets of coincident
$n-k$ and $k$ branes separated at distance $r \sim \Lambda/M^2$.
In this respect this model is analogous to the model of
\cite{ddrg}. This is unimportant for the present purposes, since we are
interested in the long distance brane interactions. The Intriligator-Thomas
type models\cite{it} with no supersymmetric ground state can also be
constructed. (See \cite{bhoo} for some brane model building in this
direction)} However, because supersymmetry is broken, the one-loop
corrections to the K{\"a}hler
metric lift the
plateau and generate the inter-brane potential
\begin{equation}
 V(r) \simeq \Lambda^4(1 + (a\lambda^2 - bg^2)){\rm ln}(Mr) 
\end{equation}
where $a$ and $b$ are positive one-loop factors and $g$ is a gauge
coupling. This potential comes from the one-loop K\"ahler renormalization
by the particles of mass $M^2r$. The heavier modes do not contribute
because of (by assumption) higher supersymmetry.
According to infrared-ultraviolet duality, the same asymptotic form should
be recovered via the tree-level closed string exchange, which is indeed
the case, since the long distance physics is dominated by the 
massless mode exchange, which give ln$(r)$ potential in the leading order.

Now the point is that $\lambda$ and $g$ should be understood as the
running functions of $\phi =M^2r$ and taking this running into account, the
potential should get a "log-corrected" form
\begin{equation}
 V(r) \simeq \Lambda^4(1 - (c_1 - c_2{\rm ln}(rM)){\rm ln}(Mr)
\end{equation}
where $c_1$ and $c_2$ are of the order of one and two-loop factors
respectively. So depending on the balance between parameters, this potential
can have a minimum at $ln(Mr) \sim c_1/c_2$, stabilizing branes at
very large distances. Of course, several things are not answered in this
toy model. For instance, what about higher loop corrections, or how
SUSY-breaking affects the open-closed
duality? However, it illustrates the main idea. Some details are given in
\cite{zura}.

\section{Attractive Branes at High Temperature}.

In high temperature supersymmetric gauge theories, the points of the
restored
gauge symmetries are always the local minima of the free energy.
Since the coincident branes correspond to the enhanced gauge symmetry
points,
it is expected that branes get stabilized on top of each other
at sufficiently high temperatures.\footnote{We will assume temperatures
below the Hagedorn point. See, e.g. \cite{hag} for some recent
discussions.}

Let us have a closer look at the nature of the zero-$T$ repulsive
potential
of the branes at $r=0$. We can do this from the point of view of an
effective field theory. In this picture relative displacement of the
branes is described by an expectation value to the scalar
field $\phi = M^2r$. In $D$-brane picture, when
branes are on top of each other
there is an enhanced gauge symmetry, which gets broken by $\phi$ VEV when
branes are separated. The gauge fields that get masses from $\phi$ are
the open string modes stretched between the different branes. The mass of the
lightest
modes from such a string will be $\sim \phi$. Now, in the unbroken
supersymmetry limit, branes are BPS states and $\phi$ is modulus
with exactly flat potential. Appearance of the inter-brane potential
in this language corresponds to the lifting of the flat moduli space
by supersymmetry breaking soft terms. In particular, the fact that
branes are repulsive means that $\phi = 0$ point became unstable
due to supersymmetry breaking, or in the other words, $\phi$ got a
negative soft mass. By dimensional grounds the curvature at $\phi = 0$
should be
decided by the supersymmetry breaking effects on the brane:
\begin{equation}
V(\phi) = -\phi^2 m_s^2
\end{equation}
where $m_s$ is the scale of supersymmetry breaking on the brane.
For large distances $\phi \gg M$ we should recover the usual inverse power
(or log$(r)$) behavior
\begin{equation}
V(\phi) = M^4(\alpha + b_i{e^{{\phi \over m_i'}} \over (\phi/M)^{N - 2}} -
{1\over (\phi/M)^{N -2}})
\end{equation}
Although this form looks singular for $\phi \rightarrow 0$, singularity
is just a reflection of the fact that some modes become massless at $\phi
=0$ and must be "integrated in". This will smooth out singularity at
the origin.
 
Now let us take into account the effect of the high temperature.
For $\phi = 0$ the string modes that get masses from its VEV are
in equilibrium and their contribution to the free energy creates a
positive $T^2$ mass term for $\phi$, so that the resulting curvature term
becomes
\begin{equation}
V(\phi)_T =  (cT^2 - m_s^2)\phi^2 +....
\end{equation}
where $c$ is a model-dependent factor. This effect will stabilize the
branes on top of each other all the way down to a certain critical
temperature $T_c \sim m_s$ below which $\phi$ gets destabilized and
branes roll away from each other. It is crucial that $T_c$ is set by $m_s$
and not by $M$. Thus the vacuum energy density of the
branes $\sim M^4$ can dominate over the thermal energy $\sim T^4$ and
trigger inflation.

\section{Thermal Brane Inflation}

The resulting inflationary scenario is quite simple. We assume that
there was a period of an early inflation with a high reheat temperature
$T_{in} \sim M$ at the end of which some of the repelling
branes appeared to sit on top of each other and got stabilized by the
thermal effects as suggested above.
Once the temperature drops below
$T_{in}$ the potential energy takes over and the
inflation results all the way until $T$ drops below $T_c$ and potential
gets destabilized. Thus, the number of available e-foldings is given by
\begin{equation}
n_e = {\rm ln}(T_{in}/T_c)
\end{equation}
Taking $T_{in} \sim 10$TeV and $T_c \sim 10^3 -10$MeV we find a maximal
possible number $n_e = 10-15$ or so. This is enough to get rid of
unwanted relics like bulk gravitons.

\section{Stretching the Superstring Dark Matter}.

 Now, let us show that in this scenario the certain amount of a superheavy
dark
matter is expected to be produced in form of the superstrings that get
stretched between the branes.\footnote{After this work was done, I learned
from K.Benakli about his work in progress on a different possibility of
using winding modes as a dark matter in high Planck scale scenario.}
 
The possibility of producing a superheavy dark matter during a 
"conventional" inflation or preheating has been discussed in the
literature\cite{superheavy}. The possibility of
heavy
particle
production during the thermal inflation\cite{thermal} is the closest
four-dimensional counterpart of our scenario.

We suggest that in the brane picture there is yet another way
of generating a superheavy dark matter in form of the sub-millimeter size
superstrings.
When branes sit on top of each other,
the lowest modes from the strings that stretch between them are effectively
massless 
and are in thermal equilibrium. Their starting number density is 
given by an initial temperature
$N_{string} \sim T_{in}^3$. We will take $T_{in} \sim M$. As soon as
temperature
drops to below $T_{in}$ the potential energy takes over and branes
inflate.
The string number density then drops exponentially fast $\sim {\rm
e}^{-3n_e}$ and is $N_{string} \sim T_c^3$ right at the end of inflation.
After this point brane
bound state gets destabilized and
branes move away stretching the strings between them. In the field theory
language this means that exited string modes are getting masses
from the $\phi$ VEV. These modes become non-relativistic and their
number density freezes out within the time $\sim m_s^{-1}$.
Let us take as an example the potential (\ref{thepotential}) with a single
repulsive
mode
of mass $m$.
Then, right after the end of inflation Universe is left with
strings of the mass
$\phi_0 \sim M^2r_0 \sim M^2/m$ and the initial number density $N_{in}
\sim T_c^3$.
The energy stored in this dark matter is $\rho_{in} \sim T_c^3M^2/m$,
which is a tiny fraction of the initial energy density of the
oscillating branes $\rho_{osc} \sim M^4$.
The brane oscillations reheat
the universe to the temperature\footnote{For the crude estimate we
will omit the model-dependent numerical factors, which otherwise may be
important, e.g. due to large
number of species, or due to loop suppression of the inflaton couplings.}
\begin{equation}
T_R \sim \sqrt{{m_{\phi}^3 \over \phi_0^2}M_P}
\label{reheat}
\end{equation}
Where $m_{\phi} \sim M(m/M)^{{N\over 2}}$ is the oscillation
frequency, the mass of the
oscillating inflaton field $\phi$.
After this point the string energy density scales as
$T^3$ so that the present day abundance can be estimated as
\begin{equation}
\Omega_{string} = \rho_{string}/\rho_c  \sim 10^{9}{\rm GeV}{T_c^3 \over
mM^2}T_R
\end{equation}
Now, from the graviton over-closure there is a strong bound on
$T_R$\cite{add1}.
which in the case of two extra dimensions gives $T_R \sim$MeV,
even for $M \sim 10$TeV.
From (\ref{reheat}) this gives $m \sim 10$MeV or so.
Taking this numbers, the right abundance
$\Omega_{string} \sim 0,3$ could have resulted if
 $T_c \sim 1{\rm GeV}$.

\section{Bulk Gravitons}

Let us now, discuss dilution of the bulk gravitons by the brane inflation.
First let us note that gravitons produced after the final reheating are safe
as
far as $T_R \sim $MeV, which in turn puts constraint on
$m$. Thus, the main problem are the gravitons produced at the initial
stage of reheating, just before the brane inflation. Let us estimate their
present abundance. Their initial number density is\cite{add1}
\begin{equation}
\rho_{gr} = {T_{in}^{N + 5} \over M^{N + 2}}M_P.
\end{equation}
This will be further reduced by a factor $\sim (T_{c}/T_{in})^3$ during
the
period of the thermal brane inflation, and  subsequently by a dilution
factor $\sim (T_R/M)^4$ during the period of the brane oscillations. After
reheating the graviton energy density scales as $T^3$ and the
condition for it to never dominate the Universe becomes
\begin{equation}
 {T_{in}^{N + 2} \over M^{N + 6}}M_PT_{c}^3T_R < 10^{-9}
{\rm GeV}
\end{equation}
Assuming $T_{in} \sim M$ this can be satisfied if $T_c \sim $MeV or so.
Note however that this inequality is very sensitive to the value of
$T_{in}$ and thus can be accommodated even for larger $T_c$ provided
$T_{in}$ is somewhat smaller than $M$.

\section{Discussions (or the Role of the Bulk Vacuum Energy)}.

We have discussed the issue of the infrared brane-brane stabilization
which may have some important consequences.

First, the large distance brane-brane stabilization, may be responsible
for the large size of the extra dimensions. For this, one has to assume
that branes "decide" the size of the bulk volume, that is (in the
leading order) the potential for the radius modulus comes purely from the
brane-brane interactions. However, when branes get
stabilized at large distances, usually, the potential is very shallow, the
mass of the inter-brane mode is $\sim 1/r_0$. So one may wonder that the
bulk vacuum energy can generate the stronger potential for $r$ that
would destroy this stabilization. For instance, a constant
$\Lambda_{bulk}$
term would generate a power-law $\sim r^n$ potential.
We want to stress that, in general, this is not necessarily the case if at
the tree-level supersymmetry is broken on the brane and gets transmitted
to the bulk only through some messengers like gravity. In such a case
there can be a zero bulk vacuum energy to "start with". For a moment let
us consider two extra dimensions (the role of these will become more clear
on an explicit example below).\footnote{We understand that
the special role of the two extra dimensions was also pointed out
in\cite{bdn},\cite{sn} but from a different perspective.}
Then, the naive dimensional analysis
suggests that the integrated bulk vacuum energy should scale as
\begin{equation}
\sim M^4f({\rm ln(rM)})
\end{equation}
where $f$ is some (non-exponential) function of ln(rM).
This naive argument relies on the fact that although the number of
available bulk modes (lighter then the cut-off scale) is enormous
$\sim (rM)^N$, the supersymmetry breaking in each multiplet is tiny
(suppressed by $\sim (rM)^{-N}$). However, the things are somewhat
more involved since the scaling law can depend how the SUSY-breaking
scales with a distance from the brane.
To illustrate the point let us consider a simple example with two
transverse dimensions, which explicitly demonstrates how the log-scaling
can appear.
As it was suggested in \cite{dshifman}, the supersymmetry breaking
on the brane can simply result from the fact that the brane is a non-BPS
soliton. The stability of the brane can be due to the topological charge.
In particular, for two extra dimensions global vortexes can serve as
non-BPS solitons\cite{dshifman}. Let us imagine a model
with two dimensions compactified on a sphere $S_2$. Let $\Phi$
be a complex scalar field defined on the sphere and we assume that
$\Phi$ has a non-zero expectation value $\Phi = v$ due to whatever
dynamics. Assume that $\Phi$ transforms under a global $U(1)$ symmetry
$\Phi \rightarrow {\rm e^{i\alpha}}\Phi$. Then, its expectation value
breaks $U(1)$ but in general leaves supersymmetry unbroken.
Supersymmetry can be maximally broken if we discuss topologically
nontrivial winding configuration (vortex) which (in a flat space limit)
asymptotically look as\cite{vilenkin}:
\begin{equation}
\Phi|_{\rho\rightarrow \infty} \rightarrow v{\rm e}^{i\theta}
\end{equation}
where $\rho$ is the distance from the core, and $\theta$ is an polar
angle. Consider a vortex-anti-vortex pair "stuck" at the opposite poles
of $S_2$. The bulk vacuum energy of the system coming from the gradient
term diverges logarithmically with vortex-anti-vortex distance $r$, which
in our case sets the size of extra dimensions:
\begin{equation}
V(r) \sim \int_{\delta}^{r} \rho d\rho |\partial_{a}\Phi|^2 \sim v^2{\rm
ln}(rM)
\end{equation}
where, $\delta$ is the size of the core. Thus, we recover the $log(r)$
behavior. Note that, if it was only the
potential energy of the vortex core, the supersymmetry would be unbroken
at the tree level in the bulk. The logarithmic behavior of the bulk
energy, can be understood as a result of a {\it tree level} transmition of
the supersymmetry breaking from the brane to the bulk by the derivatively
coupled massless Nambu-Goldstone field. The "local strength" of this
breaking ($\Lambda_{bulk}$) scales as $\sim 1/\rho^2$ as function of the
distance from brane so that integrated vacuum energy is $\sim {\rm
ln}(r)$.
 In this example, Goldstone expectation value is a part of the winding
configuration, however, in more generic case it may just play the role of
a messenger at the loop-level. The above example demonstrates that the
role of two extra dimensions can be crucial.
Note that, for the three
transverse dimensions with
point-like branes the scaling could be
different. For instance, for the global monopoles stuck on $S_3$ the
scaling would be linear in $r$.

Above discussion shows that it is not unusual for the bulk cosmological
constant to have a log-scaling behavior and may not dominate over the
direct inter-brane potential, which may be responsible for generating
the large size of extra dimensions.

 Finally, even if large inter-brane separation is not directly responsible
for the large extra dimensions, it can still have a interesting
cosmological application since can lead to a brief period of the thermal
brane inflation with an acceptably low reheating temperature and
superheavy dark matter.

%\vspace{0.5cm}

%{\bf Acknowledgments}: \hspace{0.2cm} 

\end{document}